\documentclass[12pt]{amsart}

\newcommand{\R}{\mathbb R}
\newcommand{\HH}{\mathcal H}
\newcommand{\x}{\mathbf x}
\newcommand{\p}{\mathbf p}
\newcommand{\jj}{\mathbf j}
\newcommand{\nn}{\mathbf n}
\newcommand{\vol}{\mathbf{vol}}
\newcommand{\grad}{\mathop{\rm grad}\nolimits}

\begin{document}
\title{Generalized Schrodinger equation
for free field}

\author{A. V. Stoyanovsky}

\thanks{Partially supported by the grant RFFI N~04-01-00640}

\email{stoyan@mccme.ru}

\begin{abstract}
We give a logically and mathematically self-consistent procedure of quantization of free scalar
field, including quantization on space-like surfaces. A short discussion of possible generalization
to interacting fields is added.
\end{abstract}

\maketitle

\section*{Introduction}

This is a mathematical paper, although it touches matters of quantum field theory.

The Schrodinger equation for free scalar field (see, for example, [1,2]) reads
\begin{equation}
ih\frac{\partial\Psi}{\partial t}=\int\left(-\frac{h^2}2\frac{\delta^2}{\delta u(\x)^2}+\frac12(\grad u(\x))^2
+\frac{m^2}2u(\x)^2\right)\Psi\,d\x.
\end{equation}
Here $\Psi$ is the unknown functional depending on a number $t$ and on a real function $u(\x)$, $\x=(x_1,\ldots,x_n)$;
$\frac{\delta}{\delta u(\x)}$ is the variational derivative. A known problem is to give a mathematical
sense to equation (1) and close equations and to solve them, so as to obtain the Green functions
of a free field in the answer. In other words, this is the problem of logically and mathematically self-consistent
quantization of a free field.

Traditionally one solved equation (1) in the Fock Hilbert space containing functionals of the form
$$
\Psi(u(\cdot))=\Psi_0(u(\cdot))\exp\left(-\frac1{2h}\int\hat u(\p)\hat u(-\p)\omega_\p d\p\right),
$$
where $\Psi_0$ is a polynomial functional in $u(\cdot)$;
$\p=(p_1,\ldots,p_n)$, $\hat u(\p)=\frac1{(2\pi)^{n/2}}\int e^{-i\p\x}u(\x)d\x$, $\omega_\p=\sqrt{\p^2+m^2}$.
It is easy to see that on these functionals the right hand side of equation (1) equals infinity. To
overcome this, one subtracted an ``infinite constant'' from the Hamiltonian in the right hand side of equation (1),
reducing the Hamiltonian to a normally ordered expression of creation operators
$\frac 1{\sqrt{2\omega_\p}}\left(-h\frac{\delta}{\delta\hat u(-\p)}+\omega_\p\hat u(\p)\right)$ and annihilation
operators $\frac 1{\sqrt{2\omega_\p}}\left(h\frac{\delta}{\delta\hat u(-\p)}+\omega_\p\hat u(\p)\right)$.
This procedure causes mathematical objections (aesthetical ones), as well as physical ones
(subtraction of an infinite constant is usually motivated by the fact that we are interested only in
the differences of energy levels, but here one forgets that energy is theoretically nonmeasurable at all in relativistic
quantum dynamics, and the only measurable quantities are the scattering sections).
Besides that, as shown in [3], one can find formal objections as well: namely, in the Fock space for $n>1$
one cannot give sense to the relativistically invariant generalization of equation (1) called the
Tomonaga--Schwinger equation, describing quantization on curved space-like surfaces.

In the same paper [3] it is pointed out that a non-contradictory quantization of a free field on space-like
surfaces is possible in the framework of algebraic quantum field theory. The purpose of the present paper
is to describe in detail this logically self-consistent procedure of quantization of a free field at the
mathematical level, and to discuss related questions.

In \S1, in the general framework of field theory, we derive a variational differential equation from
the variational principle, called by us the generalized Schrodinger equation
[7,8,9] and looking similar to the Tomonaga--Schwinger equation (which has not been ever given a rigorous
mathematical sense up to now). In the particular case of
flat space-like surfaces of constant time this equation reduces to the Schrodinger equation.

In \S2 we discuss the problem of giving a mathematical sense to the usual and generalized Schrodinger equation,
at least in the case of free scalar field.
This discussion continues the discussion started in the Introduction to the paper [10], but it can be read
independently. The result of this discussion is that, seemingly, the only possible way to describe dynamics
of a free quantum field is not in the space of states --- functionals
$\Psi$ as in equation (1) (the Schrodinger representation), but only in the space of observables
(the Heisenberg representation), which are elements of an abstract algebra --- an infinite dimensional
analog of the Weyl algebra. This useful argument can be applied, for example,
to two-dimensional conformal field theory: seemingly, it is more consistent to describe it
with the help of abstract fields --- observables rather than with the help of dynamics in the space of states.

\S3 is devoted to computations in the Weyl algebra.

In 3.1 we prove integrability of the generalized Schrodinger equation for a scalar field with
self-action. (In this general case the generalized Schrodinger equation in the Weyl algebra
seems to have a bad physical sense,
so we present the result as a formally mathematical one, and use it only for free fields.)

In 3.2 we present an explicit solution of the Schrodinger equation for free scalar field with a source.
In the answer we get usual Green functions of free scalar field, given by the Feynman propagator.
This computation is essentially known in quantum field theory.

Finally, \S4 is devoted to a short discussion of possible generalization of these results to the case of
interacting fields. This is a work in progress, so here the exposition is incomplete.

The author is grateful to V. V. Dolotin for useful discussions.

\section{Formal derivation of the generalized Schrodinger equation [7,8,9]}

The idea of derivation of the generalized Schrodinger equation is related with the idea of Feynman path integral.
This integral describes the imagined generalization of the wave theory to the situation of a multidimensional
variational problem. Respectively, derivation of the generalized Schrodinger equation is a direct generalization
of the derivation of the quantum mechanical Schrodinger equation to the situation of the multidimensional variational
principle.

\subsection{Formula for variation of action}
Consider the action functional of the form
\begin{equation}
J=\int_D F(x^0,\ldots,x^n,u^1,\ldots,u^m,u^1_{x^0},\ldots,u^m_{x^n})\,
dx^0\ldots dx^n,
\end{equation}
where $x^0,\ldots,x^n$ are the independent variables, $u^1,\ldots,u^m$ are the dependent
variables, $u^i_{x^j}=\frac{\partial u^i}{\partial x^j}$,
and integration goes over an $(n+1)$-dimensional surface $D$ (the graph of the functions $u^i(x)$)
with the boundary $\partial D$ in the space $\R^{m+n+1}$.

Assume that for each $n$-dimensional parameterized surface $C$
in $\R^{m+n+1}$, given by the equations
\begin{equation}
x^j=x^j(s^1,\ldots,s^n),\ \ u^i=u^i(s^1,\ldots,s^n)
\end{equation}
and sufficiently close (in $C^\infty$-topology) to a fixed $n$-dimensional surface,
there exists a unique $(n+1)$-dimensional surface $D$ with the boundary $\partial D=C$ which is an
extremal of the integral (2), i.~e. the graph of a solution to the Euler--Lagrange equations.
Denote by $S=S(C)$ the value of the integral (2) over the surface $D$.

Then one has the following well known formula for variation of the functional $S$
(for the definition of variation and variational derivative, see 2.1):
\begin{equation}
\delta S=\int_C \left(\sum p^i\delta u^i-\sum H^j\delta x^j\right)\,ds,
\end{equation}
or
\begin{equation}
\begin{aligned}{}
\frac{\delta S}{\delta u^i(s)}&=p^i(s),\\
\frac{\delta S}{\delta x^j(s)}&=-H^j(s),
\end{aligned}
\end{equation}
where
\begin{equation}
\begin{aligned}{}
p^i&=\sum_l(-1)^lF_{u^i_{x^l}}
\frac{\partial(x^0,\ldots,\widehat{x^l},\ldots,x^n)}{\partial(s^1,\ldots,s^n)},\\
H^j&=\sum_{l\ne j}(-1)^lF_{u^i_{x^l}}u^i_{x^j}
\frac{\partial(x^0,\ldots,\widehat{x^l},\ldots,x^n)}{\partial(s^1,\ldots,s^n)}\\
&+(-1)^j(F_{u^i_{x^j}}u^i_{x^j}-F)
\frac{\partial(x^0,\ldots,\widehat{x^j},\ldots,x^n)}{\partial(s^1,\ldots,s^n)}.
\end{aligned}
\end{equation}

Here
$\frac{\partial(x^1,\ldots,x^n)}{\partial(s^1,\ldots,s^n)}=\left|
\frac{\partial x^j}{\partial s^i}\right|$ is the Jacobian;
the cap over a variable means that the variable is omitted; summation
over the index $i$ repeated twice is assumed. For derivation of this formula see, for example, [8],
or [11].

Note that the coefficients before the Jacobians in the formula for $H^j$
coincide, up to sign, with the components of the energy-momentum tensor.

Note also that the quantities $p^i$ and $H^j$ depend on the numbers $u^i_{x^j}$
characterizing the tangent plane to the $(n+1)$-dimensional surface $D$.
These numbers are related by the system of equations
\begin{equation}
u^i_{x^j}x^j_{s^k}=u^i_{s^k},\ \ \ i=1,\ldots,m,\ k=1,\ldots,n.
\end{equation}
Hence, only $m(n+1)-mn=m$ numbers among $u^i_{x^j}$ are independent. Therefore
$m+n+1$ quantities $p^i$ and $H^j$ are related, in general, by $n+1$ equations. $n$ of these equations
are easy to find:
\begin{equation}
p^iu^i_{s^k}-H^jx^j_{s^k}=0,\ \ \ k=1,\ldots,n.
\end{equation}
The remaining $(n+1)$-th equation depends on the form of the function $F$. Denote it by
\begin{equation}
\HH(x^j(s),u^i(s),x^j_{s^k},u^i_{s^k},p^i(s),-H^j(s))=0.
\end{equation}

From $n+1$ equations (8) and (9) one can, in general, express the quantities $H^j$
as functions of $p^i$ (and of $x^l$, $u^i$, $x^l_{s^k}$, $u^i_{s^k}$):
\begin{equation}
H^j=H^j(x^l,u^i,x^l_{s^k},u^i_{s^k},p^i),\ \ \ j=0,\ldots,n.
\end{equation}

\subsection{The generalized Hamilton--Jacobi equation}
Substituting (5) into equations (8,9) or into equations (10), we obtain
\begin{equation}
\begin{aligned}{}
\frac{\delta S}{\delta u^i(s)}u^i_{s^k}+\frac{\delta S}{\delta x^j(s)}
x^j_{s^k}&=0,\ \ \ k=1,\ldots,n,\\
\HH\left(x^j,u^i,x^j_{s^k},u^i_{s^k},
\frac{\delta S}{\delta u^i(s)},\frac{\delta S}{\delta x^j(s)}\right)&=0,
\end{aligned}
\end{equation}
or
\begin{equation}
\frac{\delta S}{\delta x^j(s)}+H^j\left(x^l,u^i,x^l_{s^k},u^i_{s^k},
\frac{\delta S}{\delta u^i(s)}\right)=0,\ \ \ j=0,\ldots,n.
\end{equation}
The system of equations (11) or (12), relating the values of variational derivatives
of the functional $S$
at one and the same point $s$, can be naturally called the generalized Hamilton--Jacobi
equation. The first $n$ equations of the system (11) correspond to the fact that the function $S$ does
not depend on concrete parameterization of the surface $C$.

\medskip
{\bf Example (scalar field with self-action).}
Let
\begin{equation}
F(x^\mu,u,u_{x^\mu})=\frac{1}{2}(u_{x^0}^2-\sum_{j\ne0}u_{x^j}^2)-V(x,u)
=\frac12u_{x^\mu}u_{x_\mu}-V(x,u)
\end{equation}
in the standard relativistic notations, where the index $\mu$ is pushed down using the metric
$(dx^0)^2-\sum_{j\ne0}(dx^j)^2$.
A computation gives the following generalized Hamilton--Jacobi equation:
\begin{equation}
\begin{aligned}{}
x^\mu_{s^k}\frac{\delta S}{\delta x^\mu(s)}+
u_{s^k}\frac{\delta S}{\delta u(s)}&=0,\ \ k=1,\ldots,n,\\
\vol\frac{\delta S}{\delta\nn(s)}+
\frac{1}{2}\left(\frac{\delta S}{\delta u(s)}\right)^2
+\frac12\vol^2&du(s)^2+\vol^2V(x(s),u(s))=0.
\end{aligned}
\end{equation}
Here $\vol^2=D^\mu D_\mu$ is the square of the volume element on the surface,
$D^\mu=(-1)^\mu \frac{\partial(x^0,\ldots,\widehat{x^\mu},\ldots,x^n)}{\partial(s^1,\ldots,s^n)}$,
the vector $(D_\mu)=\vol\cdot\nn$ is proportional to the unit normal $\nn$ to the surface,
the number $\vol\frac{\delta S}{\delta\nn(s)}=D_\mu\frac{\delta S}{\delta x^\mu(s)}$ is proportional to
the variation $\frac{\delta S}{\delta\nn(s)}$ of the functional $S$ under the change of the surface in the
normal direction, and the number
$\vol^2du(s)^2=(D_\mu u_{x^\mu})^2-(u_{x^\mu}u_{x_\mu})(D_\nu D^\nu)$ is proportional to the scalar square
$du(s)^2$ of the differential $du(s)$ of the function $u(s)$ on the surface.
\medskip

The generalized Hamilton--Jacobi equation was written in particular cases by many authors,
see, for example, the book [12] and references therein. In [12] one can also find
a theory of integration of the generalized Hamilton--Jacobi equation in the particular case of
two dimensional variational problems, and in [8] a theory in the general case.

\subsection{Generalized canonical Hamilton equations}
Suppose that the surface $D$ is parameterized by the coordinates
$s_1$, $\ldots$, $s_n$, $t$. The generalized canonical Hamilton equations express the dependence
of the variables $p^i,u^i$ on $t$, if we assume that the dependence of $x^j$ on $(s,t)$ is given. The equations read
\begin{equation}
\begin{aligned}{}
u^i_t&=\frac{\delta}{\delta p^i(s)}\int H^jx^j_t(s')\,ds', \\
p^i_t&=-\frac{\delta}{\delta u^i(s)}\int H^jx^j_t(s')\,ds'.
\end{aligned}
\end{equation}
For their derivation, see [8]. They are equivalent to the Euler--Lagrange equations. They can be also written in the
following form:
\begin{equation}
\frac{\delta \Phi(u^i(\cdot),p^i(\cdot);x^j(\cdot))}{\delta x^j(s)}=\{\Phi,H^j(s)\},
\end{equation}
where $\Phi(u^i(\cdot),p^i(\cdot);x^j(\cdot))$ is an arbitrary functional of functions $u^i(s)$, $p^i(s)$
changing together with the surface $x^j=x^j(s)$, and
\begin{equation}
\{\Phi_1,\Phi_2\}=\sum_i\int\left(\frac{\delta \Phi_1}{\delta u^i(s)}
\frac{\delta \Phi_2}{\delta p^i(s)}-\frac{\delta \Phi_1}{\delta p^i(s)}\frac{\delta \Phi_2}{\delta u^i(s)}\right)ds
\end{equation}
is the Poisson bracket of two functionals $\Phi_l(u^i(\cdot),p^i(\cdot))$, $l=1,2$. In [8]
the generalized canonical Hamilton equations are identified with the equations of characteristics
for the generalized Hamilton--Jacobi equation.

\subsection{Generalized Schrodinger equation}
Assume that the functions $H^j$ (10) are polynomials with respect to the variables $p^i$.
Let us make the following formal substitution in the generalized Hamilton--Jacobi equation (12):
\begin{equation}
\begin{aligned}{}
&\frac{\delta S}{\delta x^j(s)} \to -ih\frac{\delta}{\delta x^j(s)},\\
&\frac{\delta S}{\delta u^{i'}(s)}=p^{i'} \to -ih\frac{\delta}{\delta u^{i'}(s)}.
\end{aligned}
\end{equation}
Here $i$ is the imaginary unit, $h$ is the Planck constant.
We obtain a system of linear variational differential equations
which can be naturally called the generalized Schrodinger equation:
\begin{equation}
-ih\frac{\delta \Psi}{\delta x^j(s)}+H^j\left(x^l,u^{i'},x^l_{s^k},u^{i'}_{s^k},
-ih\frac{\delta}{\delta u^{i'}(s)}\right)\Psi=0,\ \ \ j=0,\ldots,n.
\end{equation}
Here $\Psi=\Psi(C)$ is the unknown complex valued functional of the functions
$x^j(s),u^{i'}(s)$.

The system (19) can be written also in the form of type (11), if we assume that
the left hand side of equation (9) is polynomial with respect to the variables $H^j,p^{i'}$:
\begin{equation}
\begin{aligned}{}
&\frac{\delta \Psi}{\delta u^{i'}(s)}u^{i'}_{s^k}+\frac{\delta \Psi}{\delta x^j(s)}
x^j_{s^k}=0,\ \ \ k=1,\ldots,n,\\
&\HH\left(x^j,u^{i'},x^j_{s^k},u^{i'}_{s^k},
-ih\frac{\delta}{\delta u^{i'}(s)},-ih\frac{\delta}{\delta x^j(s)}
\right)\Psi=0.
\end{aligned}
\end{equation}
The first $n$ equations of the system (20) mean that the functional
$\Psi(C)$ does not depend on the parameterization of the surface $C$.

\medskip
{\bf Example.} In example (13) we obtain the following
generalized Schrodinger equation:
\begin{equation}
\begin{aligned}{}
&x^\mu_{s^k}\frac{\delta\Psi}{\delta x^\mu(s)}+
u_{s^k}\frac{\delta\Psi}{\delta u(s)}=0,\ \ k=1,\ldots,n,\\
ih\vol\frac{\delta\Psi}{\delta\nn(s)}&=
-\frac{h^2}{2}\frac{\delta^2\Psi}{\delta u(s)^2}
+\vol^2\left(\frac12du(s)^2+V(x(s),u(s))\right)\Psi.
\end{aligned}
\end{equation}

\subsection{Parameterization by space variables}

Since the value of the functional $\Psi(C)$ does not depend on the parameterization
of the surface $C$, we can choose a particular parameterization. Put
$s^1=x^1,\ldots,s^n=x^n$. In this parameterization the generalized Schrodinger equation
becomes a single equation, which we will write as the first equation of the system
(19) (with the number $j=0$). This equation can be easily computed from the equalities (6):
\begin{equation}
-ih\frac{\delta \Psi}{\delta x^0(\x)}+H\left(x^0(\x),\x,u^i(\x),
\frac{\partial x^0}{\partial\x},\frac{\partial u^i}{\partial\x},
-ih\frac{\delta}{\delta u^i(\x)}\right)\Psi=0,
\end{equation}
where $\x=(x^1,\ldots,x^n)$, $\frac{\partial u^i}{\partial\x}=
\left(\frac{\partial u^i}{\partial x^1},\ldots,
\frac{\partial u^i}{\partial x^n}\right)$, and
\begin{equation}
\begin{aligned}{}
H&=H\left(x^0,\x,u^i,\frac{\partial x^0}{\partial\x},
\frac{\partial u^i}{\partial\x},p^i\right)
=\sum_i p^iu^i_{x^0}-F(x^0,\x,u^i,u^i_{x^0},u^i_{x^j}),\\
u^i_{x^j}&=\frac{\partial u^i}{\partial x^j}-u^i_{x^0}
\frac{\partial x^0}{\partial x^j},\ \ \ j=1,\ldots,n,\\
p^i&=F_{u^i_{x^0}}-\sum_{j=1}^n F_{u^i_{x^j}}\frac{\partial x^0}{\partial x^j}
=\frac{\partial F}{\partial u^i_{x^0}}.
\end{aligned}
\end{equation}
Thus, $H$ is the Legendre transform of the Lagrangian $F$ with respect to the variables
$u^i_{x^0}$. Equation (22) looks approximately like the Tomonaga--Schwinger equation
[4,5,6].
For any given functional $\Psi_0(u^i(\x))$ of functions on a fixed
space-like surface $x^0=x^0(\x)$,
equation (22) describes evolution of the functional $\Psi_0$ under the change of the space-like surface.
The surface should be space-like in the case of standard Lagrangians of quantum field theory, since
otherwise the denominators of the coefficients of
equation (22) can vanish, cf. example (21) (formula (36) below).

Considering evolution of flat space-like surfaces $x^0(\x)=const=t$, we come to an evolutional equation
on a functional $\Psi(t,u^i(\x))$:
\begin{equation}
ih\frac{\partial\Psi}{\partial t}=\int H\left(t,\x,u^i(\x),
\frac{\partial u^i}{\partial\x},-ih\frac{\delta}{\delta u^i(\x)}\right)
\Psi\,d\x.
\end{equation}
This equation is called the (functional differential)
Schrodinger equation.

\section{On the mathematical sense of the usual and generalized Schrodinger equation}

In the case of free scalar field, $V(x,u)=\frac{m^2}2u^2$ in example (13),
the Schrodinger equation amounts to equation (1). At first sight it seems that this equation
can be directly given a mathematical sense. Indeed, one can give the following
(well known) rigorous definition of variational derivatives.

\subsection{Definition of variational derivatives}

Let $\Psi$ be a functional on a space of infinitely differentiable functions
$u(s)$. Let us endow this space with a structure of a complete nuclear
topological vector space or the union of such spaces.
As a rule, we will consider the Schwartz space of rapidly decreasing functions $u(s)$
without specifying it explicitly. Let us call by the weak differential or by the variation of
the functional $\Psi$ the functional
\begin{equation}
\delta\Psi(u(\cdot);\delta u(\cdot))=\lim_{\varepsilon\to0}\frac{\Psi(u(\cdot)+\varepsilon\,\delta u(\cdot))
-\Psi(u(\cdot))}{\varepsilon}.
\end{equation}
Let us call the functional $\Psi$ continuously differentiable if $\delta\Psi$ is defined and continuous
as a functional of two arguments $u(\cdot)$, $\delta u(\cdot)$, which independently run over the space of functions.
In this case $\delta\Psi$ is automatically linear with respect to the second argument, i.~e., it is a distribution as
a functional of $\delta u(\cdot)$. This distribution is denoted
$\frac{\delta\Psi}{\delta u(s)}=\frac{\delta\Psi}{\delta u(s)}(u(\cdot))$ and called
the variational derivative of the functional $\Psi$. Thus, the variational derivative is defined by the symbolic
equality
\begin{equation}
\delta\Psi=\int\frac{\delta\Psi}{\delta u(s)}\delta u(s)\,ds,
\end{equation}
reminding the definition of partial derivatives (in which $s$ takes a discrete set of values, $\int$ is replaced
by $\sum$, and $\delta$ is replaced by $d$).

Repeatedly differentiating $\delta\Psi$ with respect to $u(\cdot)$, we obtain the definition of the second
variation
$\delta^2\Psi(u(\cdot);\delta_1u(\cdot),\delta_2u(\cdot))$, the third variation, etc.
We also require their continuousness with respect to all arguments.
The second variation is symmetric and bilinear with respect to $\delta_1u(\cdot)$ and $\delta_2u(\cdot)$.
The second variational derivative $\frac{\delta^2\Psi}{\delta u(s_1)\delta u(s_2)}$ is a symmetric
distribution in $s_1$, $s_2$, defined due to the Schwartz kernel theorem by the equality
\begin{equation}
\delta^2\Psi(u(\cdot);\delta u(\cdot),\delta u(\cdot))
=\int\frac{\delta^2\Psi}{\delta u(s_1)\delta u(s_2)}\delta u(s_1)\delta u(s_2)\,ds_1ds_2.
\end{equation}
If this distribution can be restricted to the diagonal $s_1=s_2=s$, then its restriction gives
the variational derivative $\frac{\delta^2\Psi}{\delta u(s)^2}$ present in equation (1).

Analogously one defines higher variational derivatives, the Taylor series, infinitely differentiable
and analytical functionals, etc. For them one proves analogs of many main theorems
of analysis of several variables, such as the decomposition into the Taylor polynomial
with the remainder term, the implicit function theorem, etc. (see, for example,
references in [13]).

\subsection{Various orderings of operators}

Let us return to equation (1). It is easy to see that this equation, understood literally, has no
nonzero four times differentiable solutions. Indeed, consider the second derivative
$\frac{\partial^2\Psi}{\partial t^2}$. In the expression for this derivative following from (1),
we should transfer the operator $\frac{\delta^2}{\delta u(\x)^2}$ through the operator of multiplication
on $u(\x')^2$, and this would give the square of the delta function $\delta(\x-\x')^2$.

To overcome this problem, consider more general equations (19), (22), (24). In these equations
we have not yet determined the ordering of non-commuting operators $\frac{\delta}{\delta u(s)}$
and $u(s')$. Analogous problem takes place in quantum mechanics, where there are various recipes
of its solution. The simplest recipe is to put $\frac{\delta}{\delta u(s)}$ to the right of $u(s')$.
We have seen that this recipe does not suit us. But there are also other recipes, for example, the symmetric Weyl
recipe. In this recipe, for example, in our case instead of the operator $u(s)\frac{\delta}{\delta u(s')}$,
informally speaking, one gives sense to a symmetric expression
\begin{equation}
\frac12\left(u(s)\frac{\delta}{\delta u(s')}+\frac{\delta}{\delta u(s')}u(s)\right).
\end{equation}
Whereas in quantum mechanics the Weyl recipe and the simplest recipe give
the same algebra of (pseudo)differential operators (not going into details, cf., for example,
[14], \S18.5), in our case they give nonequivalent algebras.
The Weyl algebra defined below is the one that solves our problem: it makes possible to give sense to the equations
so that in the case of free field they have nontrivial physically interesting solutions (see \S3 below).

The Weyl algebra is attractable also due to the fact that it admits a compatible action of symplectic group.
This is important for us, since the evolution operators of the classical field equations from one space-like
surface to another one are canonical transformations preserving the Poisson bracket (17). This follows
from the canonical Hamilton equations (15). In the case of free field given by a quadratic Hamiltonian,
these operators are linear, i.~e., symplectic. As shown in the paper [3], these evolution operators
of the Klein--Gordon equation do not act on the Fock space. But on the Weyl algebra they do act,
which gives a solution of the problem of quantization of free field on space-like surfaces.

Let us proceed to realization of this program.

\subsection{Definition of the infinite dimensional Weyl algebra}
The Weyl algebra is constructed starting from a symplectic vector space.
Consider the Schwartz symplectic space of rapidly decreasing functions $(u^i(s),p^i(s))$ with the Poisson bracket (17).
Let us write it in the form
\begin{equation}
\{\Phi_1,\Phi_2\}=\int\sum_{i,j}\omega^{ij}\frac{\delta \Phi_1}{\delta y^i(s)}
\frac{\delta \Phi_2}{\delta y^j(s)}\,ds,
\end{equation}
where $y^i=u^i$ for $1\le i\le m$ and $y^i=p^{i-m}$ for $m+1\le i\le 2m$, and
$\omega^{ij}=\delta_{i,j-m}-\delta_{i-m,j}$.
The Weyl algebra is defined as the algebra of infinitely differentiable functionals $\Phi(u^i(\cdot),p^i(\cdot))$
with respect to the $*$-product (sometimes also called the Moyal product)
\begin{equation}
\begin{aligned}{}
&(\Phi_1*\Phi_2)(y^i(\cdot))\\
&=\left.\exp\left(-\frac{ih}2\int\sum_{i,j}\omega^{ij}
\frac{\delta}{\delta y^i(s)}\frac{\delta}{\delta z^j(s)}\,ds\right)
\Phi_1(y^i(\cdot))\Phi_2(z^i(\cdot))\right|_{z^i(\cdot)=y^i(\cdot)}.
\end{aligned}
\end{equation}
This product is not everywhere defined: for example, $u^i(s)*p^i(s)$ is undefined. We shall not go into
details of determining the multiplication domain, as well as details on defining a suitable topology on
the Weyl algebra. Note only that if all necessary integrals and series are defined and absolutely convergent,
then the $*$-product is associative. This is a formal check analogous to the finite dimensional case.
We will be interested only in some concrete computations in the Weyl algebra.

\subsection{The problem of states}

Thus, the operators in equations (19), (22), (24), and others will be understood as elements of the Weyl algebra.
And how will be understood states $\Psi$? These cannot be usual functionals of $u^i(\cdot)$, because the
Weyl algebra does not act on them. In the case of a finite dimensional symplectic vector space,
the Weyl algebra canonically acts on half-forms on the Lagrangian subspace. In coordinates
$q_1,\ldots,q_N,p_1,\ldots,p_N$ half-forms look $f(q_1,\ldots,q_N)(dq_1\ldots dq_N)^{1/2}$.
The fact that the Weyl algebra acts on half-forms, can be seen, for example, as follows.
The operator $p_iq_j$ of linear change of coordinates from the Weyl algebra acts as
$$
\frac12\left(q_j\frac{\partial}{\partial q_i}+\frac{\partial}{\partial q_i}q_j\right)
=q_j\frac{\partial}{\partial q_i}+\frac12\delta_{ij},
$$
and this is the action on half-forms.

What are half-forms on an infinite dimensional space of functions $u^i(s)$? Seemingly, one cannot say anything
definite at this point [10]. At least, half-forms cannot be constructed from finite dimensional spaces,
analogously to the construction of measures on an infinite dimensional space.

But actually, in order to obtain physically important quantities for free field, we need not states:
it suffices to use only operators, as will be shown in \S3. States are ``non-observable neither physically
nor mathematically''. Hence we will consider, instead of equations (19), (22), (24), the equations
for an element $\Phi(u^i(\cdot),p^i(\cdot);x^j(\cdot))$ of the Weyl algebra:
\begin{equation}
ih\frac{\delta\Phi}{\delta x^j(s)}=[H^j(s),\Phi],
\end{equation}
\begin{equation}
ih\frac{\delta\Phi}{\delta x^0(\x)}=[H(\x),\Phi],
\end{equation}
\begin{equation}
ih\frac{\partial\Phi}{\partial t}=\left[\int\!H(t,\x)d\x,\Phi\right],
\end{equation}
where
\begin{equation}
[\Phi_1,\Phi_2]=\Phi_1*\Phi_2-\Phi_2*\Phi_1
\end{equation}
is the commutator in the Weyl algebra. The classical limits of equations (31) are the Hamilton equations (16).

\section{Computations}

\subsection{Integrability of the generalized Schrodinger equation for scalar field with self-action}

Equation (31) describes the evolution of the functional $\Phi(u^i(\cdot),p^i(\cdot);x^j(\cdot))$ under the change
of a space-like surface $x^j=x^j(s)$. A natural self-compatibility condition is the integrability of this
equation, i.~e., the condition that the result of evolution depends only on the initial and final
space-like surfaces and not on the concrete way of evolution of the surface. Below the integrability is proved
for scalar field with self-action (example (13)).

Let us start with a general remark. Equation (31) splits into the system of equations
\begin{equation}
\sum_j ihx^j_{s^k}\frac{\delta\Phi}{\delta x^j(s)}=\left[\sum_ip^i(s)u^i_{s^k}(s),\Phi\right],\ \ 1\le k\le n,
\end{equation}
following from (8), and into equation (32). Equations (35) state that the functional $\Phi$
does not change under the change of variables $s$. The integrability of these equations and their
compatibility with equation (32) for scalar field with self-action are evident. Thus, it remains to prove
integrability of equation (32). A computation gives
\begin{equation}
H(\x)=\frac12\frac{\left(p(\x)-\sum_{j=1}^n\frac{\partial u}{\partial x^j}\frac{\partial x^0}{\partial x^j}\right)^2}
{1-\sum_{j=1}^n\left(\frac{\partial x^0}{\partial x^j}\right)^2}
+\frac12\sum_{j=1}^n\left(\frac{\partial u}{\partial x^j}\right)^2+V(x^0(\x),\x,u(\x)).
\end{equation}
By a standard argument it suffices to show that
\begin{equation}
\left[-ih\frac{\delta}{\delta x^0(\x)}+H(\x),-ih\frac{\delta}{\delta x^0(\x')}+H(\x')\right]=0.
\end{equation}
And this is a direct computation left to the reader. In fact in (37) the summand with $h$
vanishes by the equality
\begin{equation}
\frac{\delta H(\x')}{\delta x^0(\x)}-\frac{\delta H(\x)}{\delta x^0(\x')}+\{H(\x),H(\x')\}=0,
\end{equation}
which follows from integrability of the classical Hamilton equations (16), and the summands with $h^3$, etc.
are absent.

\subsection{Solution of the Schrodinger equation for free scalar field with a source}
The solution of equation (33) is given by the equality
\begin{equation}
\Phi(t_1)=U(t_0,t_1)*\Phi(t_0)*U(t_0,t_1)^{-1},
\end{equation}
where
\begin{equation}
U(t_0,t_1)=T\exp\int_{t_0}^{t_1}\!\int\frac1{ih}H(t,\x)\,dtd\x,
\end{equation}
and $T\exp\int$ means the ordered exponent (the multiplicative integral):
\begin{equation}
T\exp\int_{t_0}^{t_1}\Gamma(t)dt=1+\int\limits_{t_0<t<t_1}\Gamma(t)dt+\int\limits_{t_0<t<t'<t_1}\Gamma(t)*\Gamma(t')\,dtdt'
+\ldots.
\end{equation}
Let us first put $V(x,u)=\frac{m^2}2u^2$, i.~e., consider free scalar field.
In this case $H(t,\x)$ is a quadratic expression not depending on $t$,
hence the sign $T$ before the exponent can be omitted. By the standard properties of the Weyl algebra,
$\Phi(t_1; u(\cdot),p(\cdot))$ is obtained from $\Phi(t_0;u(\cdot),p(\cdot))$ by the following linear symplectic change
of variables:
\begin{equation}
(u(t_0,\x),p(t_0,\x)=u_t(t_0,\x))\to(u(t_1,\x),p(t_1,\x)=u_t(t_1,\x)),
\end{equation}
given by the evolution operator of the canonical Hamilton equations, i.~e., by the evolution operator
of the Klein--Gordon equation
\begin{equation}
\frac{\partial^2u}{\partial t^2}-\sum_{j=1}^n\frac{\partial^2u}{(\partial x^j)^2}+m^2u=0
\end{equation}
from the Cauchy surface $t=t_0$ to the Cauchy surface $t=t_1$. Here it is rather essential that the evolution
operator is a continuous linear invertible operator in the Schwartz space of functions $(u(\x),p(\x))$.
The analogous statement is true for the evolution of the functional $\Phi$ between any two space-like Cauchy surfaces.
(For non-quadratic Hamiltonians and nonlinear classical evolution operators the analogous statement is false.)

Hence we can identify the Weyl algebras related to different space-like surfaces, by means of the evolution
operators of the Klein--Gordon equation. In other words, we can consider the Weyl algebra
$W_0$ of the symplectic vector space of solutions
$u(t,\x)$ of the Klein--Gordon equation in the whole space-time.
The symplectic form on this vector space is given by taking the Cauchy data on any space-like
surface. Below we will fix this identification of Weyl algebras of various space-like surfaces.

Now consider free scalar field with a source, i.~e., put $V(t,\x,u)=\frac{m^2}2u^2+\jj(t,\x)u$,
where $\jj(t,\x)$ is a function with compact support (a source).
Denote the corresponding element (40) of the Weyl algebra by $U_\jj(t_0,t_1)$, and the Hamiltonian by $H_\jj$,
to show the dependence on the source. Then, if the support of the function
$\jj(t,\x)$ is situated between the planes $t=t_{\min}$ and $t=t_{\max}$, then
the element
\begin{equation}
R_\jj(t_0)=U_0(t_{\max},t_0)*U_\jj(t_{\min},t_{\max})*U_0(t_0,t_{\min})
\end{equation}
of the Weyl algebra does not depend on $t_{\min},t_{\max}$. Besides that,
\begin{equation}
R_\jj(t_1)=U_0(t_0,t_1)*R_\jj(t_0)*U_0(t_0,t_1)^{-1}.
\end{equation}
Hence the element $R_\jj(t_0)=R_\jj(t_0;u(\cdot),p(\cdot))$ correctly defines an element of the Weyl algebra
of any space-like surface under our identification, i.~e., an element $R_\jj$ of the Weyl algebra $W_0$.
This element equals
\begin{equation}
R_\jj=R_\jj(u(\cdot,\cdot))=T\exp\int\limits_{-\infty}^{\infty}\!\int\frac1{ih}\jj(t,\x)u(t,\x)\,dtd\x,
\end{equation}
where
\begin{equation}
u(t,\x)=\exp\left(-\frac{t-t_0}{ih}H_0\right)*u(\x)*\exp\left(\frac{t-t_0}{ih}H_0\right)
\end{equation}
is understood as a functional on the space of solutions $u(\cdot,\cdot)$ of the Klein--Gordon equation,
i.~e., as an element of the algebra $W_0$. Let us call the element $R_\jj$ by the {\it generating functional
of operator Green functions of a free field}. Let us also call
the coefficients of the Taylor series of the functional $R_\jj$ with respect to $\jj$ at the point $\jj\equiv0$,
\begin{equation}
(ih)^N\left.\frac{\delta^NR_\jj}{\delta\jj(t_1,\x_1)\ldots\delta\jj(t_N,\x_N)}\right|_{\jj\equiv0}
=Tu(t_1,\x_1)*\ldots*u(t_N,\x_N),
\end{equation}
by the operator Green functions of a free field; here
the symbol $T$ denotes the $*$-product ordered by increasing of the variables $t_i$. The operator Green functions
are distributions in $(t_1,\x_1)$, $\ldots$, $(t_N,\x_N)$ with values in $W_0$, which are symmetric
with respect to permutation of indices.

Let us now pass to the scalar Green functions. To this end, define a linear functional on the
algebra $W_0$, called the vacuum average of an element $\Phi$ from $W_0$ and denoted
$\langle\Phi\rangle$ or $\langle0|\Phi|0\rangle$,
in the following way. The momentum representation
\begin{equation}
\hat u(p_0,\ldots,p_n)=\frac1{(2\pi)^{(n+1)/2}}\int e^{-i\sum p_jx^j}u(x^0,\ldots,x^n)\,dx
\end{equation}
of a solution $u(t,\x)$ of the Klein--Gordon equation, where $t=x^0$, is a distribution supported on two
branches of the mass surface $p_0=\pm\sqrt{\p^2+m^2}$, where $\p=(p_1,\ldots,p_n)$, $m>0$. Therefore
$u(t,\x)$ can be uniquely decomposed into the sum
\begin{equation}
u(t,\x)=u_+(t,\x)+u_-(t,\x)
\end{equation}
of a (complex) positive frequency solution $u_+(t,\x)$, whose Fourier transform is supported on the branch $p_0>0$,
and a negative frequency solution $u_-(t,\x)$, whose Fourier transform is supported on the branch $p_0<0$.
We have
\begin{equation}
\begin{aligned}{}
[u_+(t_1,\x_1),u_+(t_2,\x_2)]&=[u_-(t_1,\x_1),u_-(t_2,\x_2)]=0,\\
[\hat u_-(p),\hat u_+(p')]&=-h\delta(p+p')\delta(p^2-m^2),
\end{aligned}
\end{equation}
where $p^2=p_0^2-\sum_{j=1}^np_j^2$.
Define $\langle\Phi\rangle$ as the unique (not everywhere defined) functional with the following properties:
\begin{equation}
\langle\Phi*u_-(t,\x)\rangle=\langle u_+(t,\x)*\Phi\rangle=0,\ \ \langle1\rangle=1.
\end{equation}
Define the Green functions by the equalities
\begin{equation}
\langle u(t_1,\x_1)\ldots u(t_N,\x_N)\rangle=\langle Tu(t_1,\x_1)*\ldots*u(t_N,\x_N)\rangle,
\end{equation}
and their generating functional by the equality
\begin{equation}
Z(\jj)=\langle R_\jj\rangle.
\end{equation}
A computation left to the reader shows that the two-point Green function turns out to coincide with
the Feynman propagator
\begin{equation}
\langle u(t,\x)u(t',\x')\rangle\ \widehat{ }\ =ih\frac{\delta(p+p')}{p^2-m^2+i\varepsilon},
\end{equation}
and the generating functional of the Green functions is given by the usual expression
\begin{equation}
Z(\jj)=\exp\frac{-i}{2h}\int\frac{\hat\jj(p)\hat\jj(-p)}{p^2-m^2+i\varepsilon}\,dp.
\end{equation}
One can define, in a standard way, the Fock space linearly generated by the vectors
\begin{equation}
|p_{(1)},\ldots,p_{(N)}\rangle=\hat u_+(p_{(1)})\ldots\hat u_+(p_{(N)})|0\rangle,
\end{equation}
and formally assign an operator in the Fock space to an element $\Phi$ of the Weyl algebra $W_0$, with the matrix elements
\begin{equation}
\langle0|\hat u_-(-p'_{(1)})\ldots\hat u_-(-p'_{(N')})*\Phi*\hat u_+(p_{(1)})\ldots\hat u_+(p_{(N)})|0\rangle.
\end{equation}
But for many important operators responsible for local dynamics, for example, for the Hamiltonian $\Phi=H_0$,
the expression (58) is undefined.

\section{Discussion}

The question arises how to generalize the above mathematical construction of the Green functions to the case of
interacting fields, for example, to the scalar field with $\varphi^4$-self-action, $V=\frac{m^2}2u^2+\frac g{4!}u^4$
in Example (13) above. In this case, the evolution operators of the classical Hamilton equations (15) are
nonlinear canonical transformations of the phase space of the field, from one space-like surface to another.
We could apply the above procedure of computation of the Green functions to the naive Heisenberg equation (31)
in the Weyl algebra. As shown in 3.1, this equation is integrable.
When computing the vacuum average $\langle\,\rangle$, we can replace the
coupling constant $g$ by a function $g(t,\x)$ with compact support, as in Bogolubov's approach. Then for the
space-like surfaces not intersecting the support of the function $g(t,\x)$, the field is free, and we can compute
the vacuum average using the rules (52) above. But seemingly this procedure gives infinite result,
the integrals in question being divergent in the same manner as the usual Feynman diagram integrals.

This is not surprising, because we have chosen the Weyl algebra with the Moyal product for all space-like
surfaces, which is unnatural due to non-linearity of classical evolution transformations. Instead it seems
natural to choose, instead of the Weyl algebra, an appropriate deformation of the Poisson algebra of functionals on
the phase space. This deformation could be constructed from the bundle of Weyl algebras corresponding to the
tangent space to the phase space at each point. The finite dimensional theory of such ``deformation quantization''
of symplectic manifolds is developed in Fedosov's paper [15]. The corresponding equations of field could look
as follows. Consider functionals $\Phi(x^j(s);u^i(s),p^i(s);v^i(s),r^i(s))$. Introduce the structure of
algebra on these functionals by the Moyal product (30), in which one has to put
$y^i=v^i$ for $1\le i\le m$ and $y^i=r^{i-m}$ for $m+1\le i\le 2m$. The equations have the form
\begin{equation}
\left\{
\begin{array}{l}
ih\frac{\delta\Phi}{\delta x^j(s)}=[H^j(s),\Phi],\\
ih\frac{\delta\Phi}{\delta u^i(s)}=[D_1^i(s),\Phi],\\
ih\frac{\delta\Phi}{\delta p^i(s)}=[D_2^i(s),\Phi],
\end{array}
\right.
\end{equation}
where $H^j$ are appropriate Hamiltonians, and $D_k^i$ is a connection form, an ``Abelian connection''
in Fedosov's terminology, which can originate from a symplectic connection on the phase space.
Of course, if we follow Fedosov's construction literally, then we obtain only formal series in $h$,
but the renormalized Green functions give also only formal series in $h$.

The solutions to equations (59) form an algebra of functionals $\Phi$, which gives the required
deformation of the Poisson algebra of functionals on the phase space after the substitution $v^i=r^i=0$.

One can simplify the system (59) very much if one applies the statement, proved by Fedosov in the finite
dimensional case, that any two Abelian connections are related by conjugation by an appropriate exponent
$\exp(-i\HH/h)$. If we assume that the same is true in the infinite dimensional case, then we can rewrite
the field equations (for flat space-like surfaces $t=const$) in the following simple form:
\begin{equation}
ih\frac{\partial\Phi}{\partial t}=\left[H+\HH_1,\Phi\right],
\end{equation}
where $\Phi(t;u^i(s)+v^i(s);p^i(s)+r^i(s))$ is the unknown functional, the commutator is
taken by the same rules as above (with respect to the variables $v^i$ and $r^i$), $H$ is the Hamiltonian
written in the variables $v^i$ and $r^i$, and $\HH_1$ is certain ``renormalization'' Hamiltonian. One could
hope to obtain the renormalized Green functions or something like this in the answer, but at present the
author does not know how to do that.


\begin{thebibliography}{99}
\bibitem{1} I. M. Gelfand, A. M. Yaglom, Integration in functional spaces and its applications in quantum physics,
Uspekhi Mat. Nauk (Russian Math. Surveys), 1956, vol.~11, No.~1, 77--114.
\bibitem{2} V. P. Maslov, O. Yu. Shvedov, Method of complex germ in the many-particle problem and in quantum field
theory, Moscow, Editorial URSS, 2000 (in Russian).
\bibitem{3} C.~G.~Torre, M.~Varadarajan, Functional evolution of free quantum fields,
Class. Quant. Grav. 16 (1999) 2651--2668, hep-th/9811222.
\bibitem{4} S.~Tomonaga, Prog. Theor. Phys. {\bf 1}, 27 (1946).
\bibitem{5} J. Schwinger, Phys. Rev. {\bf 74}, 1439 (1948).
\bibitem{6} P. A. M. Dirac, Lectures on quantum mechanics, Belfer Graduate School of Science, Yeshiva University,
New York, 1964.
\bibitem{7} A. V. Stoyanovsky, Analogs of the Hamilton--Jacobi and Schrodinger equations for multidimensional
variational problems of field theory, math-ph/0208034, in: Trends in Mathematical Physics Research,
C.~V.~Benton ed., Nova Science Publishers, N.Y., 2004, p.~205--208.
\bibitem{8} A. V. Stoyanovsky, Excitations propagating along surfaces, math-ph/0301036.
\bibitem{9} A. V. Stoyanovsky, Generalized Schrodinger equation and constructions of quantum field theory,
math-ph/0301037.
\bibitem{10} A. V. Stoyanovsky, Gaussian transform of the Weil representation, math-ph/0601029.
\bibitem{11} I. M. Gelfand, S. V. Fomin, Variational calculus, Fizmatlit, Moscow, 1961 (in Russian).
\bibitem{12} P. L\'evy, Probl\`emes concrets d'analyse fonctionnelle, Paris, 1951.
\bibitem{13} A. Pressley, G. Segal, Loop groups, Clarendon Press, Oxford, 1988.
\bibitem{14} L. Hormander, The analysis of linear partial differential operators, vol.~3,
Springer Verlag, 1985.
\bibitem{15} B. V. Fedosov, A simple geometrical construction of deformation quantization,
J.~Diff. Geom. 40 (1994) 213--238.
\end{thebibliography}
\end{document}